\documentclass{appolb}

\usepackage{graphicx}
\usepackage{wrapfig}
\usepackage{amsmath}
\usepackage{amssymb}
\usepackage{color}
\usepackage{nicefrac}
\usepackage{subfig}
\usepackage{cite}

\begin{document}
\title{
\vspace{-5.0cm}
\begin{flushright}
		{\small {\bf SMU-HEP-12-21}}   \\
\end{flushright}
\vspace{0.5cm}
Strange Quark PDF Uncertainty and its Implications for W/Z Production at the LHC%
\thanks{Presented by A.~Kusina at {\em Light Cone 2012}, 8-13 July 2012.}%
}
\author{A.~Kusina$^a$, T.~Stavreva$^b$, S. Berge$^c$, F.~I.~Olness$^a$, I.~Schienbein$^b$, K.~Kova\v{r}\'{\i}k$^d$, T.~Je\v{z}o$^b$,
J.~Y.~Yu$^{a,b}$, K.~Park$^a$
\address{$^a$ Southern Methodist University, Dallas, TX 75275, USA}
\address{$^b$ Laboratoire de Physique Subatomique et de Cosmologie, Universit\'e
Joseph Fourier/CNRS-IN2P3/INPG,
 53 Avenue des Martyrs, 38026 Grenoble, France}
\address{$^c$ Institute for Physics (WA THEP), Johannes Gutenberg-Universit\"at,
D-55099 Mainz, Germany}
\address{$^d$ Institute for Theoretical Physics, Karlsruhe Institute of Technology, 
Karlsruhe, D-76128, Germany}
}
\maketitle
\begin{abstract}
We investigate the impact of parton distribution functions (PDFs)
uncertainties on $W/Z$ production at the LHC,
concentrating on the strange quark PDF.
Additionally we
examine the extent to which precise measurements at the LHC
can provide additional information on the proton flavor structure.
\end{abstract}

\section{Introduction}

Parton distribution functions provide the essential link between
the theoretically calculated partonic cross-sections, and the experimentally
measured physical cross-sections involving hadrons and mesons. This
link is crucial if we are to make incisive tests of the Standard Model (SM),
and search for subtle deviations which might signal new physics.

We show that despite recent advances in the precision data
and theoretical predictions%
~\cite{Abramowicz:1984yk,Berge:1987zw,Bazarko:1994tt,Kretzer:2003it,Tzanov:2005kr},
the relative uncertainty on the heavier
flavors remains large. We will focus on the strange quark and show
the impact of these uncertainties on
the production of $W/Z$ bosons at the LHC

This work is based on Ref.~\cite{Kusina:2012vh}, and further
details can be found therein.

\section{Extracting the Strange Quark PDF}

In previous global analyses, the predominant information on the strange
quark PDF $s(x)$ came from the difference of (large) inclusive cross
sections for Neutral Current (NC) and Charged Current (CC) DIS.
For example, at leading-order (LO)
the difference between the NC and CC DIS $F_{2}$ structure
function is proportional to the strange PDF.
Because the strange distributions are small compared to the
up and down PDFs, the $s(x)$ extracted from this measurement has
large uncertainties. Lacking better information, it was commonly assumed
the distribution was of the form
\begin{equation}
s(x) = \bar{s}(x) \sim \kappa[\bar{u}(x)+\bar{d}(x)]/2
\label{eq:kappa}
\end{equation}
with $\kappa\sim\nicefrac{1}{2}$.
This approach was used, for example, in the CTEQ6.1
PDFs~\cite{Stump:2003yu}, but it does not reflect the true
uncertainty of $s(x)$, in fact
it reflects the uncertainty on the
up and down sea which is well constrained by DIS measurements.

Beginning with CTEQ6.6 PDFs~\cite{Nadolsky:2008zw} the neutrino--nucleon
dimuon data was included in the global fits to more directly constrain
the strange quark;%
   \footnote{For a detailed review of experimental constraints of
   strange PDF see ref.~\cite{Kusina:2012vh}.}
thus, Eq.~\eqref{eq:kappa} was not used, and
two additional fitting parameters were introduced to allow the strange
quark to vary independently of the $u$ and $d$ sea.
\begin{wrapfigure}{l}{0.4\textwidth}
\centering
\includegraphics[width=0.35\textwidth]{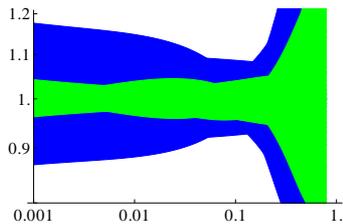}
\caption{Relative uncertainty of the $s$ PDF as a function
of $x$ for $Q=2$~GeV. The inner band is for CTEQ6.1, the
outer for CTEQ6.6 PDF set.
}
\label{fig:sband}
\end{wrapfigure}
In Fig.~\ref{fig:sband} we show the relative uncertainty bands for the
CTEQ6.6 PDF error sets (outer, blue), and for CTEQ6.1 sets (inner, green).
We observe that in case of CTEQ6.6 the relative error on the strange
quark is much larger than for the CTEQ6.1 set, particularly for $x<0.01$,
where the neutrino--nucleon dimuon data do not provide any constraints.
We expect this is a more accurate representation of the true uncertainty.

This general behavior is also exhibited in other global PDF sets with
errors~\cite{Martin:2009iq,Ball:2010de,Alekhin:2009ni,JimenezDelgado:2008hf}.
Thus, there is general agreement that the strange quark PDF is poorly
constrained, particularly in the small $x$ region.

\section{Implications for Drell-Yan W/Z Production at the LHC}

\begin{figure*}[tp]
\begin{centering}
\includegraphics[width=0.35\textwidth]{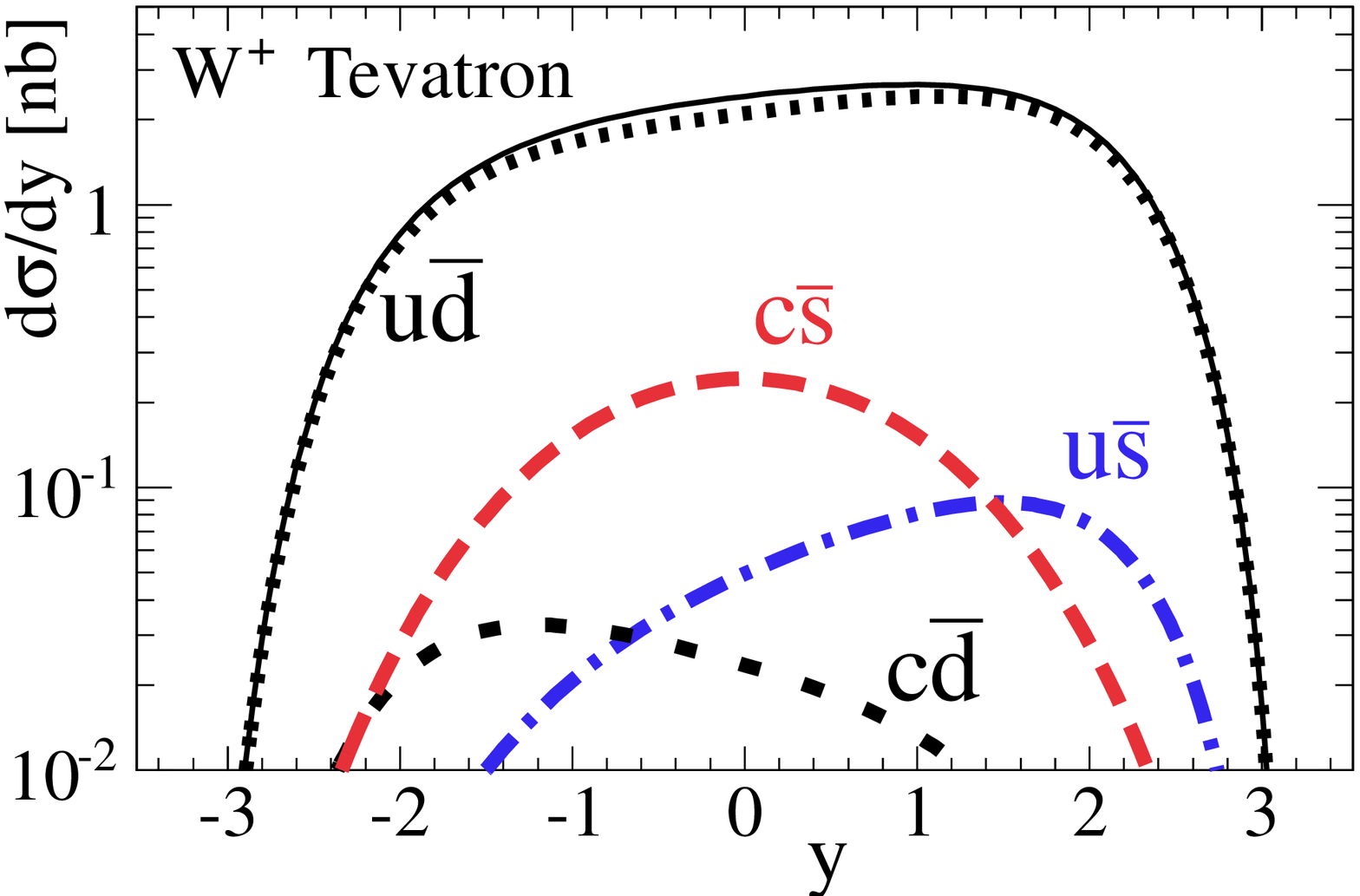}
\quad{}
\includegraphics[width=0.35\textwidth]{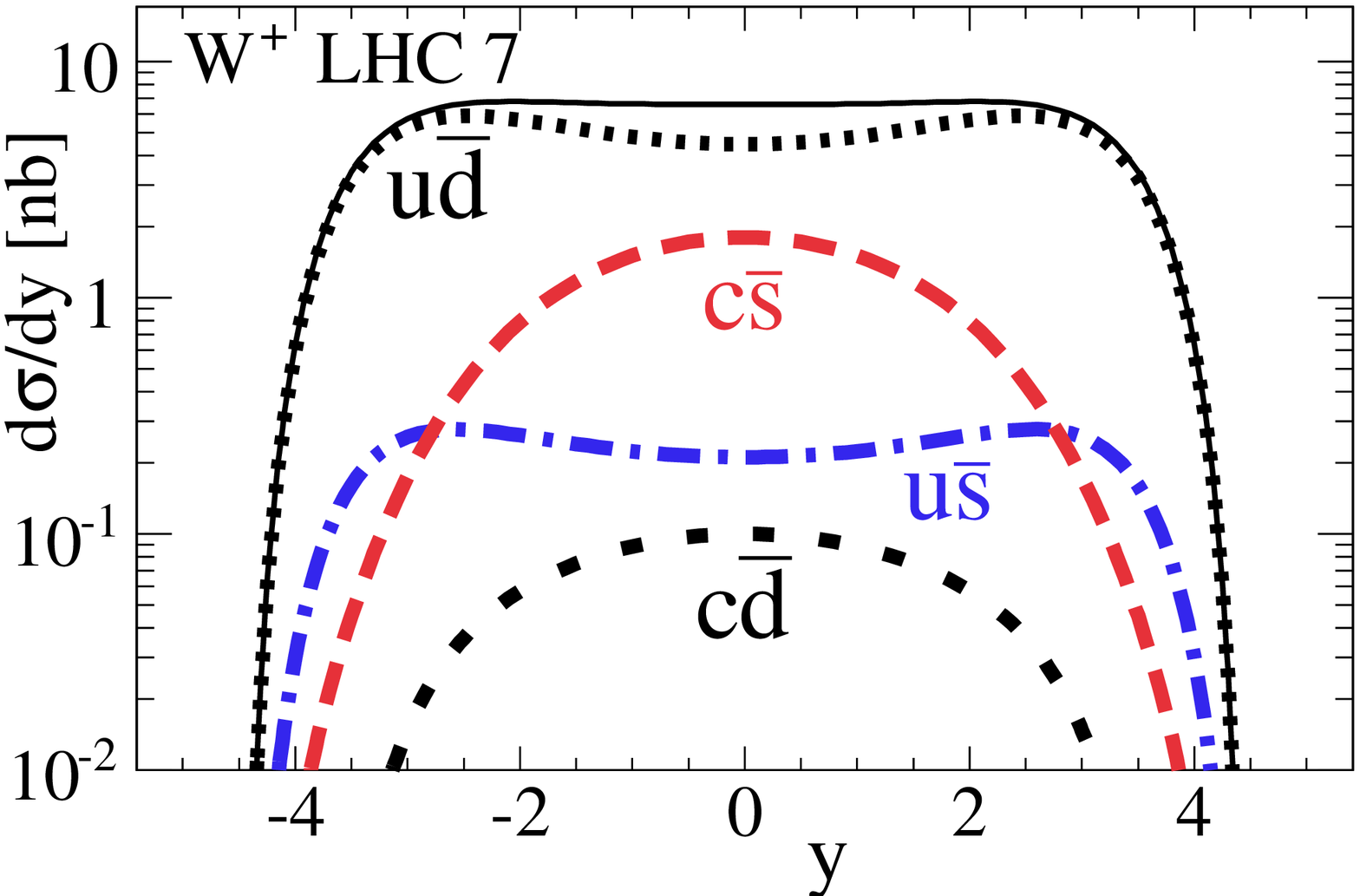}
\caption{Partonic contributions to
$d\sigma/dy$ for $W^{+}$ boson production
at LO at the Tevatron
(left), and LHC with $\sqrt{S}=7$ TeV (right).
}
\label{fig:LOlum}
\end{centering}
\end{figure*}

The Drell-Yan production of $W^{\pm}$ and $Z$ bosons at hadron colliders
can provide precise measurements for electroweak observables,
which can measure fundamental parameters of the SM.
Furthermore, the $W/Z$ boson cross section ``benchmark''
processes are intended to be used for detector calibration and luminosity
monitoring;
to perform these tasks it is essential
that we know the impact of the PDF uncertainties on these measurements.
In the following, we will investigate the influence of the PDFs on
the rapidity distributions of the Drell-Yan production process.

\subsection{Strange Contribution to $W/Z$ Production}

%
\begin{figure*}
\includegraphics[width=0.3\textwidth]{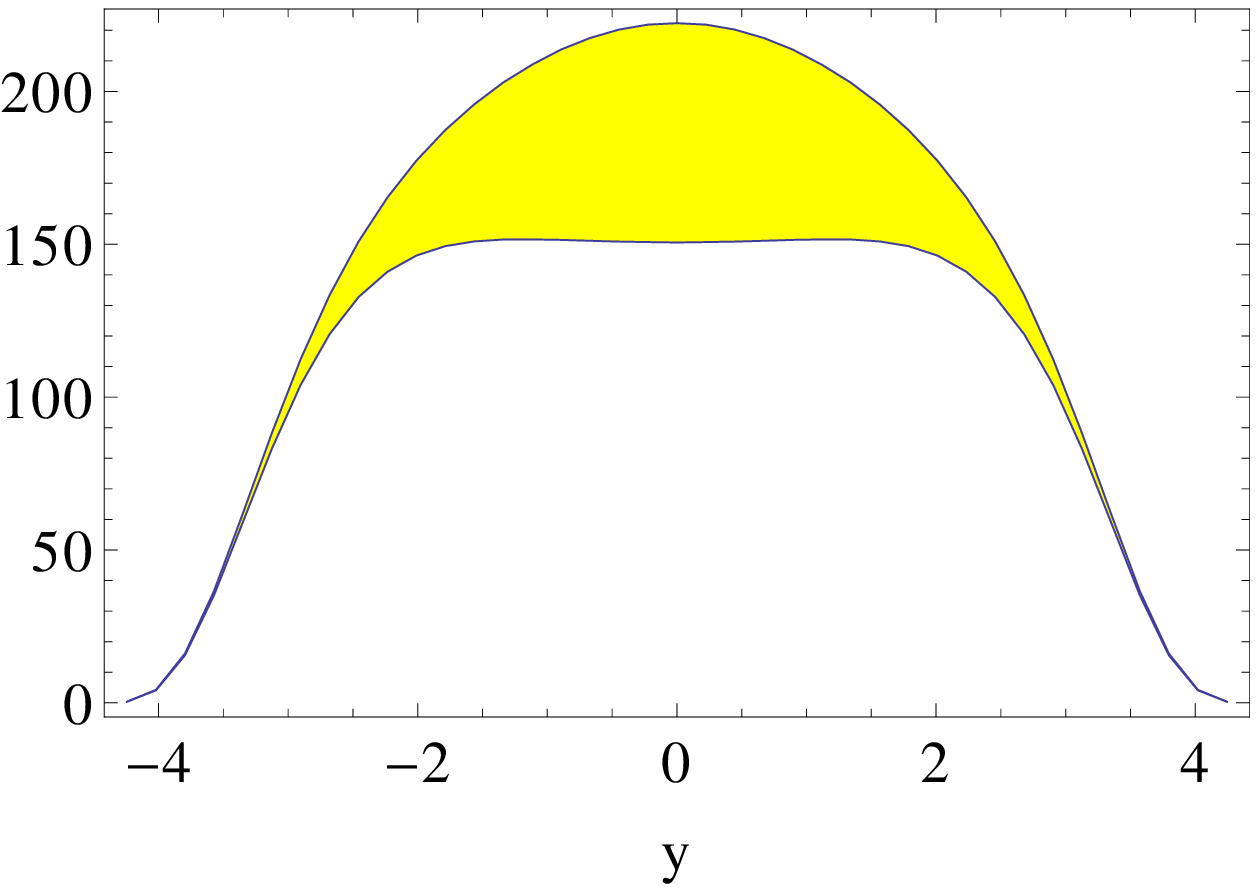}
\quad{}
\includegraphics[width=0.3\textwidth]{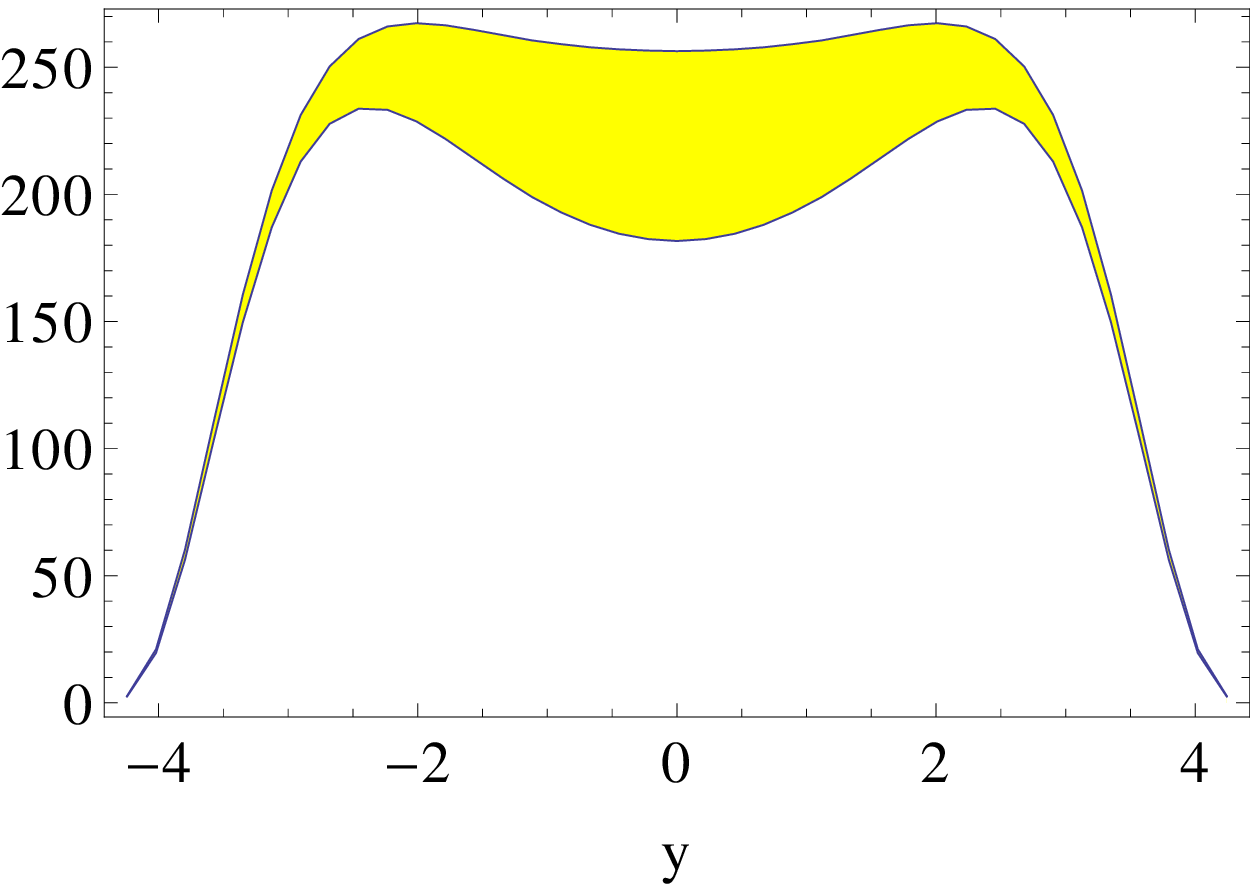}
\quad{}
\includegraphics[width=0.293\textwidth]{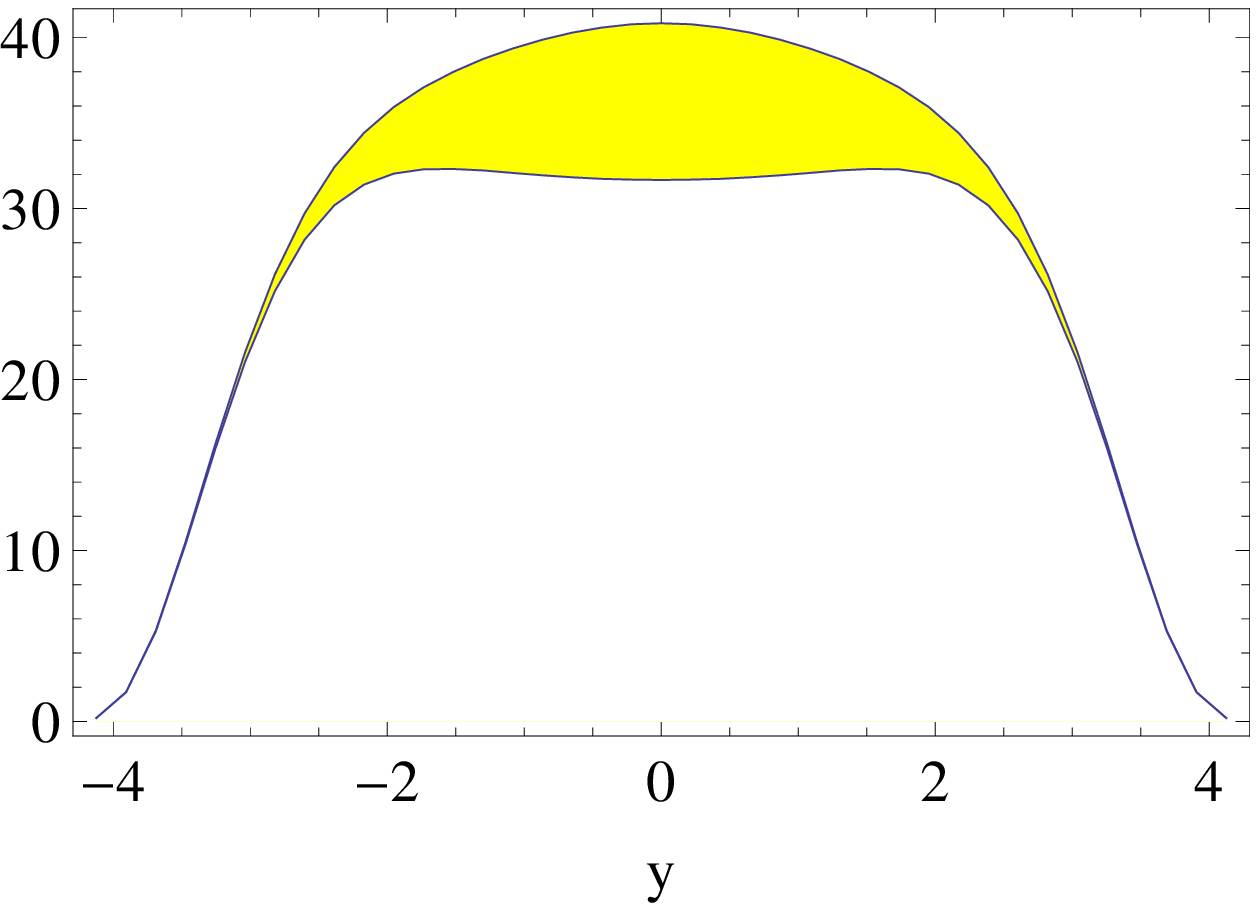}
\caption{$d^{2}\sigma/dM/dy$ in pb/GeV for $pp\to W^{-}+X$ (left),
$pp\to W^{+}+X$ (middle) and $pp\to Z,\gamma^{*}+X$ (right)
production at the LHC for 7 TeV.
\label{fig:nnlo}}
\end{figure*}

Because the LHC has a different initial state and a
higher CMS energy than the Tevatron, the relative contributions of
the partonic subprocesses of the $W/Z$ production change significantly.
At the LHC, the contributions of the second generation quarks $\{s,c\}$
are greatly enhanced.
In Fig.~\ref{fig:LOlum}
we display how the different partonic cross sections
contribute to $W^{+}$ production at LO.
In Tevatron the $u\bar{d}$ channel
contributes $90\%$ of the cross section,
while contributions from strange quarks
are comparably small $\sim9\%$.
At the LHC, subprocesses containing strange quarks are considerably
more important. The $\bar{s}u$ channel
contributes only $2\%$, while the $\bar{s}c$ channel yields $21\%$.

We also note the LHC explores a much larger rapidity range.
For channels containing strange quarks,
$|y_{W/Z}|$ can be measured up to 4.5
at the LHC, compared to 2.5
at the Tevatron; therefore
smaller values of $x$ of the strange
quark distribution can be probed.

While the LO illustration of Fig.~\ref{fig:LOlum} provides a useful
guide, in Fig.~\ref{fig:nnlo} we display the strange quark contribution
to the differential cross section $d^{2}\sigma/dM/dy$ of on-shell
$W^{\pm}/Z$ production computed at NNLO using~\cite{Anastasiou:2003ds}.
The (yellow) band represents the strange-quark initiated contributions.

The figures impressively highlight the large contribution of the
$s$ and $\bar{s}$
quark subprocesses at the LHC. Consequently it is
essential to constrain the strange PDF if we are to make accurate
predictions and to perform precision measurements.
Figure~\ref{fig:nnlo} also shows
the very different rapidity profiles of
the strange quark and valence quark terms.
This property is most evident for the case of $W^{+}$ production.
Here, the dominant $u\bar{d}$ contribution has a twin-peak structure,
while the $c\bar{s}$ distribution
has a single-peak centered at $y=0$. The total distribution is then
a linear combination of the twin-peak and single-peak distributions,
and these are weighted by the corresponding PDF. 

Therefore, a detailed measurement of the rapidity distribution of
the $W^{\pm}/Z$ bosons can yield information about the contributions
of the $s$ quark relative to the $u,d$ quarks. As this is a relative
measurement it
is reasonable to expect that this could be achieved with high precision
once sufficient statistics are collected. Consequently, this is an
ideal measurement where the LHC data could lead to stronger constraints
on the PDFs.

Note, that ATLAS has already used $W/Z$ production to infer constraints
on the strange quark distribution, and they measured $r_{s}=0.5(s+\bar{s})/\bar{d}=1.00_{-0.28}^{+0.25}$
at $Q^{2}=1.9$~GeV$^{2}$ and $x=0.023$~\cite{Aad:2012sb}.

\subsection{PDF Uncertainty of the $W/Z$ rapidity distributions \label{sub:PDFuncertainty}}

\begin{figure*}
\subfloat[]{
\label{fig:dy}
\includegraphics[width=0.3\textwidth]{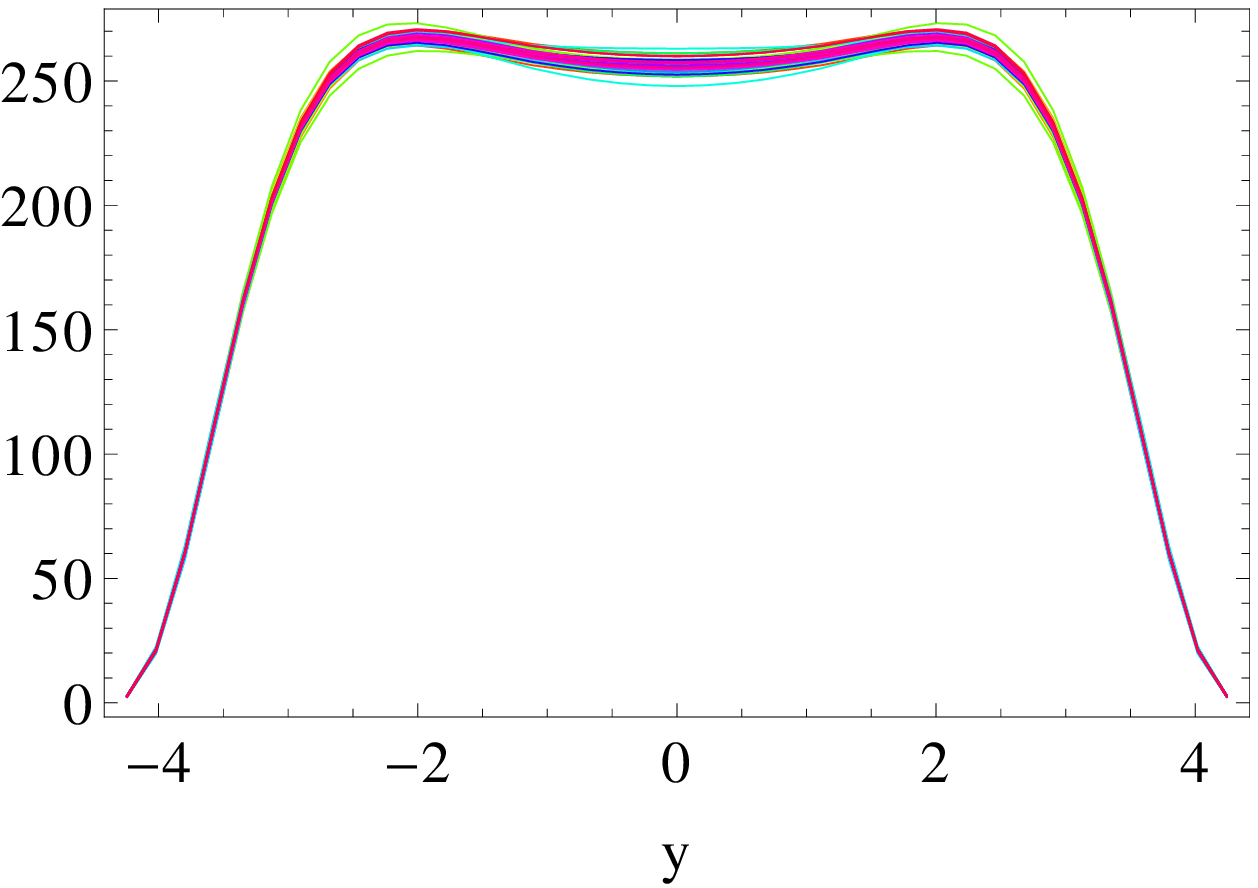}
}
\subfloat[]{
\label{fig:dy_ratio}
\includegraphics[width=0.3\textwidth]{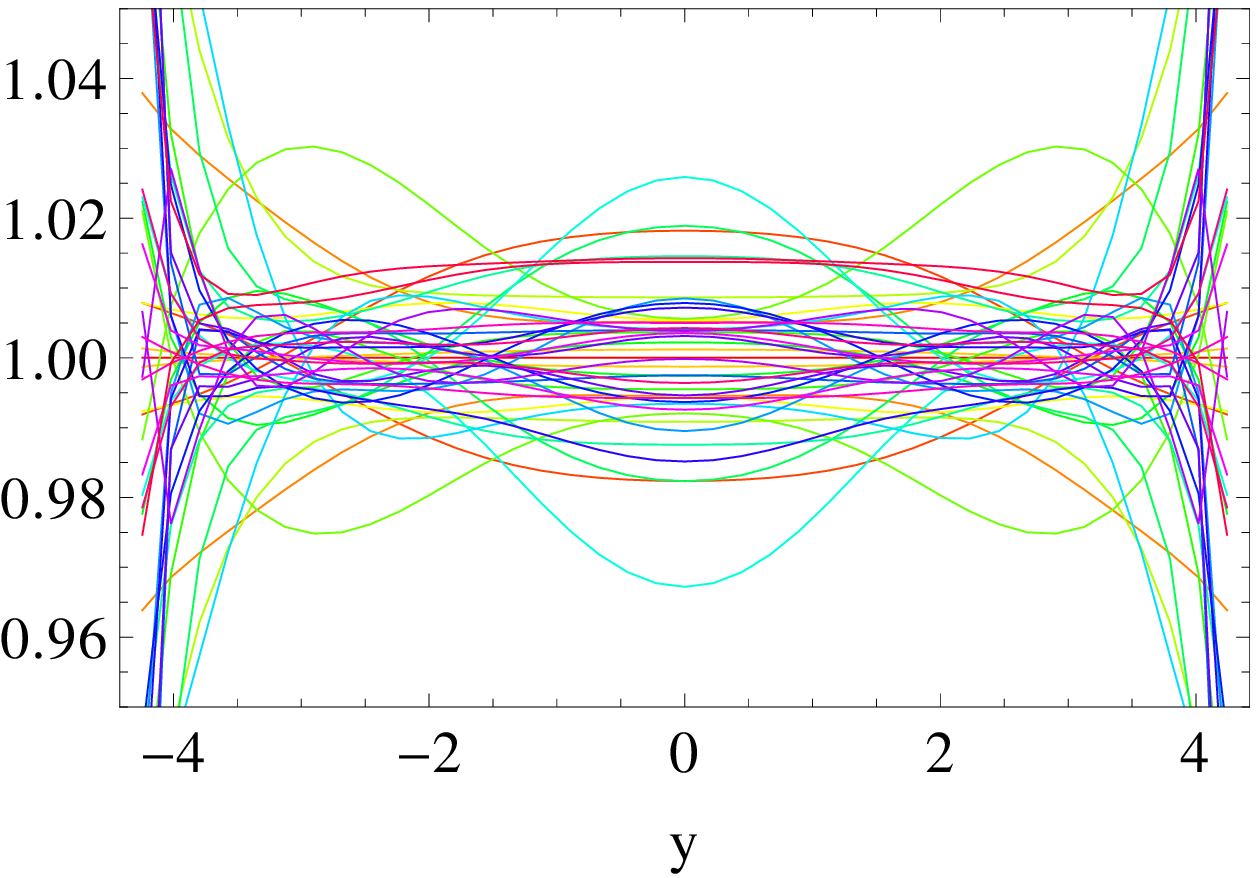}
}
\subfloat[]{
\label{fig:dy_ratio_other}
\includegraphics[width=0.3\textwidth]{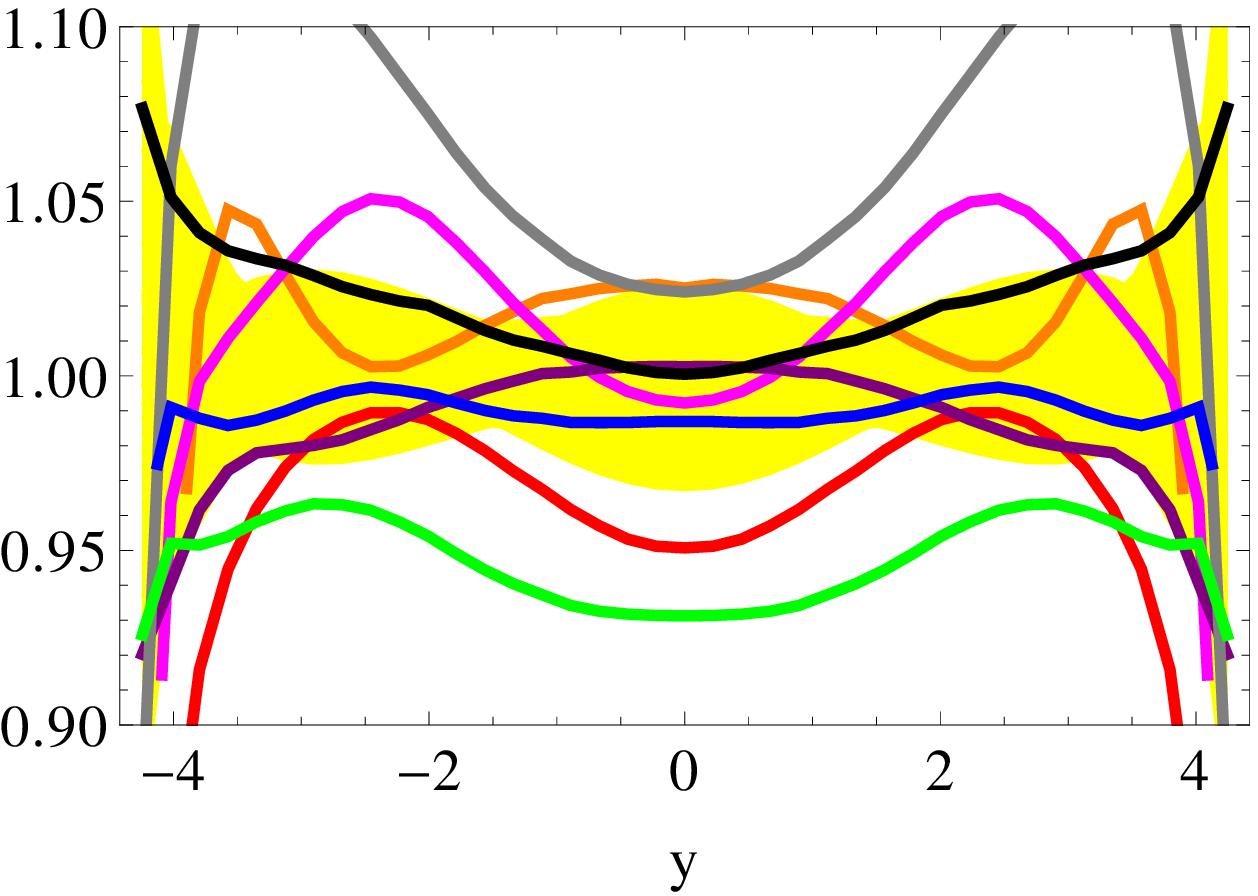}
}
\caption{
(a) $d^{2}\sigma/dM/dy$ for $pp\to W^{+}+X$ production at the LHC
with CTEQ6.6 PDF set.
(b) Fig.~(a) scaled by the central value.
(c) Yellow band representing $d^{2}\sigma/dM/dy$ for $W^{+}$ production
with CTEQ6.6 set, compared with different PDF sets,
all scaled by the central value of CTEQ6.6 set.}
\label{fig:dy_all}
\end{figure*}

In Fig.~\ref{fig:dy}, we display the differential cross section
$d^{2}\sigma/dM/dy$ for $W^{+}$ production at the LHC
($\sqrt{S}=7$~TeV) using the 44 CTEQ6.6 error PDFs
calculated at NNLO~\cite{Anastasiou:2003ds}. To better
resolve these PDF uncertainties, in Fig.~\ref{fig:dy_ratio}
we plot the ratio of the differential
cross section $d^{2}\sigma/dM/dy$ compared to the central value.
We observe that the uncertainty due
to the PDFs as measured by this band is about $\pm3\%$
for central boson rapidities.

For comparison, in Fig.~\ref{fig:dy_ratio_other} we display the (yellow)
band of CTEQ6.6 error PDFs of Fig.~\ref{fig:dy_ratio} together with
the results using a selection of contemporary PDFs, all scaled by
the central value of CTEQ6.6 set.%
   \footnote{For details on the used PDFs see ref.~\cite{Kusina:2012vh}.}
We observe that the choice of PDF sets can result in differences
ranging to about $\pm8\%$, which is well beyond the $\sim\pm3\%$
displayed in Fig.~\ref{fig:dy_ratio}.

While the band of error PDFs provides an efficient method to quantify
the uncertainty, the range spanned by the different PDF sets illustrates
that there are other important factors which should  be considered.

\subsection{Correlations of the $W/Z$ rapidity distributions }

%
\begin{figure*}[t]
\begin{centering}
\subfloat[]{
\label{fig:DoubleLHC7_65}
\includegraphics[width=0.35\textwidth]{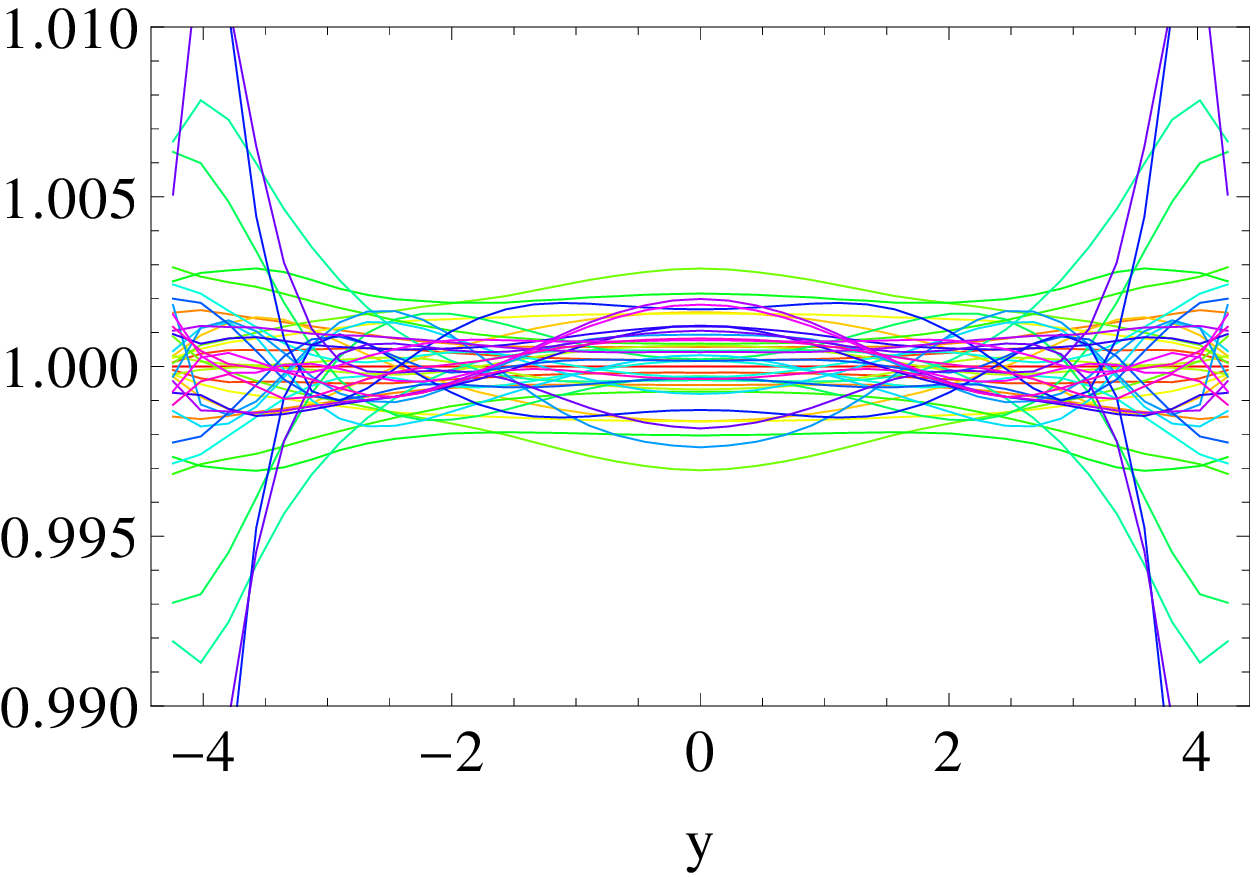}
}
\quad{}
\subfloat[]{
\label{fig:DoubleLHC7_66}
\includegraphics[width=0.35\textwidth]{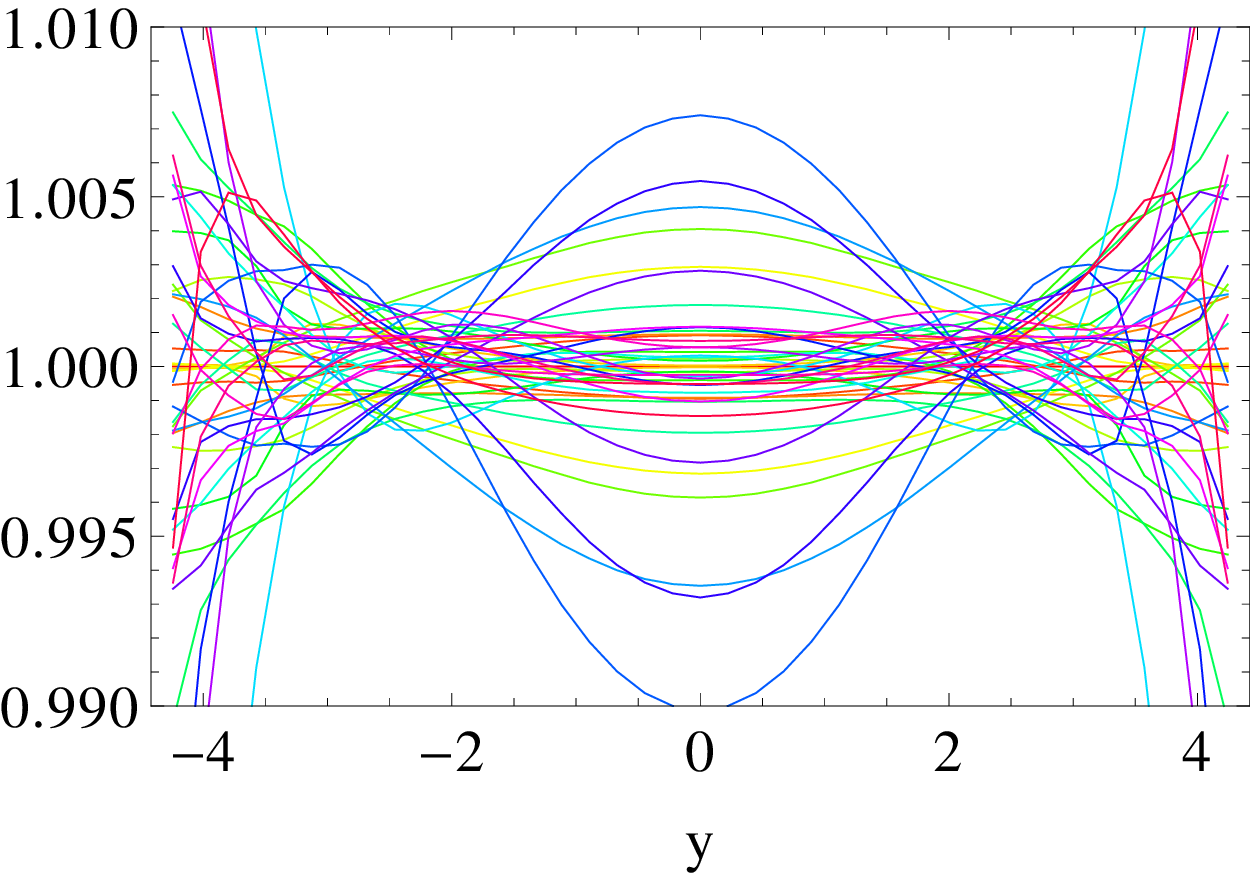}
}
\caption{
Double ratio $R$ as defined in Eq.~(\ref{eq:Ratio_RWZ}) for the
LHC with CTEQ6.5 (a) and CTEQ6.6 (b) calculated at NNLO.
\label{fig:DoubleLHC7}}
\end{centering}
\end{figure*}

The leptonic decay modes of the $W/Z$ bosons provide a powerful tool
for precision measurements of electroweak parameters such as the $W$
boson mass. As the leptonic decay of the $W$ boson contains a neutrino
($W\to\ell\nu$), the $Z$ boson production process
($Z\to\ell^{+}\ell^{-}$) is used to calibrate the leptonic $W$ process.
This method works to the extent
the production processes of the $W$ and $Z$ bosons are correlated.

One possible measure to gauge the correlation of the PDF uncertainty
is the ratio of the sum of the differential $W^{+}$ and $W^{-}$
cross sections with respect to the differential $Z$ boson cross
section, normalized to the distribution of the central PDF set.
We define:
\begin{equation}
R = \left[\frac{d\sigma(W^{+}+W^{-})}{d\sigma(Z)}\right]
   /\left[\frac{d\sigma(W^{+}+W^{-})}{d\sigma(Z)}\right]_{0}.
\label{eq:Ratio_RWZ}
\end{equation}
In Fig.~\ref{fig:DoubleLHC7} we plot $R$ for the CTEQ6.5 and
CTEQ6.6 PDF sets.

We observe that the uncertainty band in double ratios of
Fig.~\ref{fig:DoubleLHC7} are much smaller than for single ratio of
Fig.~\ref{fig:dy_all}, which reflects the fact that $W$ and $Z$
processes are correlated.
We also observe that the band for CTEQ6.5
(Fig.~\ref{fig:DoubleLHC7_65}) is much smaller than for
CTEQ6.6 (Fig.~\ref{fig:DoubleLHC7_66}), meaning that correlation
between $W$ and $Z$ processes is bigger when we use CTEQ6.5 PDFs.

The primary difference that is driving this result is the different
strange PDF. For CTEQ6.5 the strange quark was defined by Eq.~\eqref{eq:kappa}
while CTEQ6.6 introduced two extra fitting parameters which allowed
the strange PDF to vary independently from the up and down sea. Thus,
the uncertainty of the CTEQ6.6 distributions more accurately reflects
the true uncertainty.

\section{Conclusion}

We have investigated the constraints of the strange
PDFs and their impact on the $W/Z$ boson production at
the LHC.

We observe that the strange quark is rather poorly constrained,
particularly in the low $x$ region which is sensitive to $W/Z$ production
at the LHC. Improved analyses from neutrino DIS measurements could
help reduce this uncertainty. Conversely, precision measurements of
$W/Z$ production at the LHC may provide input to the global PDF analyses
which could further constrain these distributions.

\section*{Acknowledgments}
This work was partially supported by the U.S. Department of Energy
under grant DE-FG02-04ER41299, and the Lightner-Sams Foundation.

\bibliographystyle{utphys_spires}
\bibliography{bibStrange}

\providecommand{\href}[2]{#2}\begingroup\begin{thebibliography}{10}

\bibitem{Abramowicz:1984yk}
H.~Abramowicz {\em et al.}, {\em Z. Phys.} {\bf C25} (1984)
29.

\bibitem{Berge:1987zw}
J.~P. Berge {\em et al.}, {\em Z. Phys.} {\bf C35} (1987)
443.

\bibitem{Bazarko:1994tt}
{CCFR Collaboration} Collaboration, A.~Bazarko {\em et al.}, {\em Z.Phys.} {\bf
  C65} (1995) 189--198,
\href{http://www.arXiv.org/abs/hep-ex/9406007}{{\tt hep-ex/9406007}}.

\bibitem{Kretzer:2003it}
S.~Kretzer, H.~L. Lai, F.~I. Olness, and W.~K. Tung, {\em Phys. Rev.} {\bf D69}
  (2004) 114005,
\href{http://www.arXiv.org/abs/hep-ph/0307022}{{\tt hep-ph/0307022}}.

\bibitem{Tzanov:2005kr}
{NuTeV} Collaboration, M.~Tzanov {\em et al.}, {\em Phys. Rev.} {\bf D74}
  (2006) 012008,
\href{http://www.arXiv.org/abs/hep-ex/0509010}{{\tt hep-ex/0509010}}.

\bibitem{Kusina:2012vh}
A.~Kusina, T.~Stavreva, S.~Berge, F.~Olness, I.~Schienbein, {\em et al.}, {\em
  Phys.Rev.} {\bf D85} (2012) 094028,
\href{http://www.arXiv.org/abs/1203.1290}{{\tt 1203.1290}}.

\bibitem{Stump:2003yu}
D.~Stump, J.~Huston, J.~Pumplin, W.-K. Tung, H.~Lai, {\em et al.}, {\em JHEP}
  {\bf 0310} (2003) 046,
\href{http://www.arXiv.org/abs/hep-ph/0303013}{{\tt hep-ph/0303013}}.

\bibitem{Nadolsky:2008zw}
P.~M. Nadolsky, H.-L. Lai, Q.-H. Cao, J.~Huston, J.~Pumplin, {\em et al.}, {\em
  Phys.Rev.} {\bf D78} (2008) 013004,
\href{http://www.arXiv.org/abs/0802.0007}{{\tt 0802.0007}}.

\bibitem{Martin:2009iq}
A.~Martin, W.~Stirling, R.~Thorne, and G.~Watt, {\em Eur.Phys.J.} {\bf C63}
  (2009) 189--285,
\href{http://www.arXiv.org/abs/0901.0002}{{\tt 0901.0002}}.

\bibitem{Ball:2010de}
R.~D. Ball {\em et al.}, {\em Nucl. Phys.} {\bf B838} (2010) 136--206,
\href{http://www.arXiv.org/abs/1002.4407}{{\tt 1002.4407}}.

\bibitem{Alekhin:2009ni}
S.~Alekhin, J.~Blumlein, S.~Klein, and S.~Moch, {\em Phys.Rev.} {\bf D81}
  (2010) 014032,
\href{http://www.arXiv.org/abs/0908.2766}{{\tt 0908.2766}}.

\bibitem{JimenezDelgado:2008hf}
P.~Jimenez-Delgado and E.~Reya, {\em Phys.Rev.} {\bf D79} (2009) 074023,
\href{http://www.arXiv.org/abs/0810.4274}{{\tt 0810.4274}}.

\bibitem{Anastasiou:2003ds}
C.~Anastasiou, L.~J. Dixon, K.~Melnikov, and F.~Petriello, {\em Phys.Rev.} {\bf
  D69} (2004) 094008,
\href{http://www.arXiv.org/abs/hep-ph/0312266}{{\tt hep-ph/0312266}}.

\bibitem{Aad:2012sb}
{ATLAS} Collaboration, G.~Aad {\em et al.}, {\em Phys.Rev.Lett.} {\bf 109}
  (2012) 012001,
\href{http://www.arXiv.org/abs/1203.4051}{{\tt 1203.4051}}.

\end{thebibliography}\endgroup

\end{document}